\documentclass[%
reprint,
superscriptaddress,
 amsmath,amssymb,
 aps,
 pre,
]{revtex4-1}

\usepackage{graphicx}
\graphicspath{{figures/}}
\usepackage{amsmath,amssymb,amsfonts,float}
\usepackage{subfigure}
\usepackage[maxfloats=100]{morefloats}
\usepackage{dcolumn}
\usepackage{bm}
\usepackage{url}
\usepackage{algorithm}
\usepackage{algpseudocode}
\usepackage{mathtools}
\usepackage{comment}
\usepackage{multirow}
\usepackage{listings}
\usepackage{hyperref}

\newcommand{\mywd}{1.0\textwidth}

\begin{document}


\title{A non-intrusive reduced order modeling framework for quasi-geostrophic turbulence}

\author{Sk. M. Rahman}
\author{S. Pawar}

\author{O. San}
\email{osan@okstate.edu}%
\affiliation{%
 School of Mechanical and Aerospace Engineering, Oklahoma State University, Stillwater, OK 74078, USA\\
}%


\author{A. Rasheed}
\affiliation{
Department of Engineering Cybernetics, Norwegian University of Science and Technology, N-7465, Trondheim, Norway\\
}%

\author{T. Iliescu}
\affiliation{
Department of Mathematics, Virginia Tech, Blacksburg, VA 24061, USA\\
}%


\date{\today}

\begin{abstract}
In this study, we present a non-intrusive reduced order modeling (ROM) framework for large-scale quasi-stationary systems. The framework proposed herein exploits the time series prediction capability of long short-term memory (LSTM) recurrent neural network architecture such that: (i) in the training phase, the LSTM model is trained on the modal coefficients extracted from the high-resolution data snapshots using proper orthogonal decomposition (POD) transform, and (ii) in the testing phase, the trained model predicts the modal coefficients for the total time recursively based on the initial time history. Hence, no prior information about the underlying governing equations is required to generate the ROM. To illustrate the predictive performance of the proposed framework, the mean flow fields and time series response of the field values are reconstructed from the predicted modal coefficients by using an inverse POD transform. As a representative benchmark test case, we consider a two-dimensional quasi-geostrophic (QG) ocean circulation model which, in general, displays an enormous range of fluctuating spatial and temporal scales. We first illustrate that the conventional Galerkin projection based reduced order modeling of such systems requires a high number of POD modes to obtain a stable flow physics. In addition, ROM-GP does not seem to capture the intermittent bursts appearing in the dynamics of the first few most energetic modes. However, the proposed non-intrusive ROM framework based on LSTM (ROM-LSTM) yields a stable solution even for a small number of POD modes. We also observe that the ROM-LSTM model is able to capture quasi-periodic intermittent bursts accurately, and yields a stable and accurate mean flow dynamics using the time history of a few previous time states, denoted as the lookback time-window in this paper. Throughout the paper, we demonstrate our findings in terms of time series evolution of the field values and mean flow patterns, which suggest that the proposed fully non-intrusive ROM framework is robust and capable of predicting noisy nonlinear fluid flows in an extremely efficient way compared to the conventional projection based ROM framework.


\end{abstract}

\maketitle


\section{Introduction}
\label{sec:intro}
Large-scale turbulent flows, such as atmospheric and geophysical flows, are nonlinear dynamical systems which exhibit an enormous range of complex, coherent spatio-temporal scales. Over the past half century, computational approaches have made a significant contribution to characterize and understand the behavior of such flow phenomena. To resolve physical problems with high spatio-temporal variabilities through numerical simulation, one needs a high-fidelity modeling technique like direct numerical simulation (DNS). However, a huge amount of computational resources are required to capture the fine details of the flow dynamics which can become inefficient and unmanageable after some level of accuracy. Although there has been a continuous growth in computer power and performance following Moore's law during the past few decades \cite{mack2011fifty}, the progress has started to stagnate in the recent years \cite{powell2008quantum,waldrop2016chips,theis2017end,kumar2018end}. As a result, one of the most active research areas in modeling of turbulent flow dynamics is the development of efficient and robust algorithms that aim at achieving the maximum attainable quality of numerical simulations with optimal computational costs. Indeed, computational costs can be reduced by using low-fidelity models such as large eddy simulation (LES) \cite{sagaut2006large} and Reynolds-averaged Navier-Stokes (RANS) \cite{tennekes1972first} with additional approximations in the governing equations to neglect some of the physical aspects for closure modeling. Even so, these techniques require parameter calibration to approximate the true solution to any degree of confidence and may thus increase costs related to model validation, benchmark data generation, and efficient analysis of the generated data sets. As an alternative to the existing techniques, the reduced order modeling (ROM) approach has quickly become a promising approach to reduce the computational burden of high-fidelity simulations. In general, ROM works in such a way that the high-dimensional complex dynamical systems will be represented with much lower-dimensional (but dense) systems while keeping the solution quality within the acceptable range \cite{quarteroni2015reduced,taira2017modal}. An introduction to ROM methodologies can be found in recent review articles \cite{benner2015survey,taira2017modal,rowley2017model}.

There have been a significant number of strategies proposed over the years to obtain ROMs of nonlinear dynamical systems. These ROM techniques have been utilized for a wide variety of applications related to, e.g., flow control \cite{noack2011reduced,brunton2015closed,ito1998reduced}, data assimilation in weather and climate modeling \cite{daescu2008dual,cao2007reduced}, and uncertainty quantification \cite{ullmann2014pod,chen2015new,haasdonk2013reduced}. Among the different variants of ROM strategies, the Galerkin projection combined with proper orthogonal decomposition (POD) based ROMs (ROM-GP) have been utilized extensively in various areas \cite{carlberg2011efficient,lucia2004reduced,bistrian2015improved,brenner2012reduced,burkardt2006pod,iollo2000stability,freno2014proper}. POD, also known as principal component analysis (PCA), is a mathematical technique to extract the dominant statistical characteristics from turbulent flow fields by identifying the most energetic modes \cite{lumley1967structure,antoulas2001survey,chinesta2011short,benner2015survey,rowley2017model}. These few POD modes possess the fine-scale details of the system and have the capability of representing the true physics accurately. Over the years, considerable work has been done to improve the regular POD approaches \cite{hesthaven2015certified,lassila2013model,quarteroni2015reduced,berkooz1993proper,taira2017modal,aubry1992spatio,benner2015survey,rowley2004model,noack2011reduced,sieber2016spectral}.

In general, POD uses the data obtained from experiments or high-fidelity numerical simulations and generates an orthonormal set of spatial basis vectors describing the main directions (modes) by which the flow is represented optimally, in an $L_2$ sense \cite{berkooz1993proper}. The most energetic modes are kept to generate the reduced order system while the other modes are truncated. However, it has been observed that the discarded modes often contribute to the evolving dynamics of large-scale complex turbulent flow systems, like the geophysical flows \cite{lassila2014model}, resulting in instabilities and modeling errors in the solution \cite{akhtar2009stability,rempfer2000low,rowley2006dynamics,iollo2000stability}. Thus, several research efforts have been devoted to improve the stability of ROM-GP frameworks by addressing the truncated modes contributions \cite{aubry1988dynamics,rempfer1994evolution,rempfer1994dynamics,cazemier1998proper,xie2017approximate}. \citeauthor{noack2003hierarchy} \cite{noack2003hierarchy} proposed an extra `shift-mode' for accurate representation of the unstable steady solution. Several closure modeling ideas are devised to resolve the weak dissipation associated with POD modes by introducing eddy-viscosity terms (similar to LES eddy-viscosity models) \cite{sirisup2004spectral,cordier2013identification,osth2014need}. \citeauthor{san2014proper} \cite{san2014proper} improved the ROM performance by finding an optimal value for eddy-viscosity parameter with the assumption that the amount of dissipation is not identical for all the POD modes. In our recent work, we proposed an automated approach to find the eddy-viscosity parameter dynamically to stabilize the ROM-GP model \cite{rahman2019dynamic}. An alternative approach to find the eddy-viscosity parameter dynamically has been proposed by using an extreme learning machine architecture \cite{san2018extreme}. With the growing interest in data-driven modeling of ROMs using machine learning (ML) architectures, there has been another dimension of research introduced to the community for the improvement in ROM performance, referred as hybrid ROM approach. Generally, the hybridization is done by combining an imperfect physics-based model with a data-driven technique to get a hybrid scheme, and it is observed that the hybrid model shows better predictive performance than the component models \cite{rahman2018hybrid,wan2018data,xie2018data,pathak2018hybrid}.

In this paper, we develop a fully non-intrusive ROM approach as a potential alternative to already existing ROM methodologies. Indeed, physics-based (intrusive) ROM frameworks require an approximation of stabilization or regularization parameters and depend on underlying governing equation to get the solution. On the other hand, the hybrid approaches require computation of both intrusive and non-intrusive contributions, which can make the overall computation expensive. However, it is well-known that a non-intrusive approach can make the framework greatly efficient when it can be implemented successfully. With the abundance of massive amounts of data resources from high-fidelity simulations, field measurements, and experiments, the data-driven modeling approaches are currently considered some of the most promising methods across different scientific and research communities. In the past few years, artificial neural networks (ANNs) and other ML techniques have started a revolution in turbulence modeling community \cite{raissi2019physics,maulik2017neural,raissi2018multistep,lee1997application,maulik2018data,faller1997unsteady,san2018neural,wang2019non,moosavi2015efficient,kani2017dr}. Interested readers are directed to Refs.~\cite{brunton2019machine,kutz2017deep,durbin2018some,duraisamy2018turbulence,gamboa2017deep} for more on the influence of ML on fluid mechanics, specifically turbulence modeling.

With a goal to develop an efficient and robust non-intrusive ROM framework for large-scale quasi-stationary systems like geophysical flows, we propose a methodology based on long short-term memory (LSTM) recurrent neural networks. Since reduced order modeling of such noisy large-scale systems is comparatively difficult due to instabilities, which results in using a very large number of POD modes to capture the true physics, our main motivation in this study is to utilize the time series prediction capability of LSTM \cite{gers2002applying,wang2018model,yeo2019deep,sak2014long,gamboa2017deep} to capture the flow physics with a very few POD modes. As detailed in Ref.~\cite{yeo2019deep}, LSTM is very robust in predicting a very noisy sequential time series. In general, for this type of random time series, LSTM does the prediction using its own internal dynamics, which is found stable and close to the true solution \cite{yeo2019deep,mohan2018deep}. For this reason, we choose to utilize LSTM architecture based on our problem of interest, which is the large-scale quasi-stationary turbulence. However, we emphasize that this non-intrusive model can be developed by using other relevant neural network architectures as well. We also mention that the development of ROM using POD and LSTM has been used only in a few other works and proven to be successful in capturing the temporal dynamics of fluid flows. \citeauthor{wang2018model} \cite{wang2018model} proposed a non-intrusive ROM (NIROM) based on LSTM and used it to predict laminar flows.
In another recent work, \citeauthor{vlachas2018data} \cite{vlachas2018data} proposed a data-driven method based on LSTM to predict the state derivative of chaotic systems using the short-term history of the reduced order states. The predicted derivatives are then used for one-step forward prediction of the high-dimensional dynamics. The authors further developed a hybrid framework combining mean stochastic model and LSTM for data-driven to extend the forecasting capability of the proposed approach. To do the dimensionality reduction, the authors utilized discrete Fourier transform, singular value decomposition, and empirical orthogonal functions. \citeauthor{mohan2018deep} \cite{mohan2018deep} developed a non-intrusive ROM using LSTM and POD for flow control applications through a detailed analysis on different ROM-LSTM training and testing hyper-parameter tuning parameters. Even though the authors' idea of developing non-intrusive ROM based LSTM by replacing Galerkin projection is similar to our present work, their work is mostly focused on exploring the capability of LSTM in modeling the flow in reduced order space for data sets with less randomness. Indeed, the data sets with less randomness have more ``memory" in it, i.e., there are persistent or anti-persistent trends and thus, are more controllable through LSTM hyper-parameters. On the other hand, our present work is focused on exploring the capability of ROM-LSTM framework in resolving large-scale geophysical flow problem where the data sets mostly do not follow any particular trend.
To this end, we develop a modular ROM-LSTM approach in chaotic and quasi-stationary systems to see whether it can overcome the instability issues associated with conventional ROMs for noisy dynamical systems.
To assess our proposed framework, we consider the barotropic vorticity equation (BVE) representing the single-layer quasi-geostrophic (QG) model as an example of the quasi-stationary system. We observe a remarkably efficient predictive performance by the proposed framework based on LSTM (ROM-LSTM) through a number of numerical experiments and analyses.

The layout of the paper is as follows: Section~\ref{sec:qg} provides an overview of the barotropic vorticity equation describing a single-layer QG ocean model. In Section~\ref{sec:romgp}, dimension reduction through Galerkin-projection and proper orthogonal decomposition is illustrated briefly. Our proposed non-intrusive ROM-LSTM framework with a brief introduction to LSTM are presented in Section~\ref{sec:romlstm}. In Section~\ref{sec:results}, we evaluate the predictive performance of the proposed ROM framework with respect to the standard ROM and full order model solutions. Finally, Section~\ref{sec:con} provides a summary of this study and the conclusions drawn from it.

\section{Single-layer quasi-geostrophic (QG) Ocean Circulation Model}
\label{sec:qg}
In the present study, we consider the simple single-layer QG ocean circulation model to develop and assess the performance of different ROM approaches. Following Refs.~\cite{holm2003modeling,san2011approximate}, we consider the single-layer QG problem as a benchmark for wind-driven, large-scale oceanic flow. Wind-driven flows of mid-latitude ocean basins have been studied frequently by modelers using idealized single- and double-gyre wind forcing, which helps in understanding various aspects of ocean dynamics, including the role of mesoscale eddies and their effect on mean circulation. However, modeling the vast range of spatio-temporal scales of the oceanic flows with all the relevant physics has always been challenging. As a result, the numerical simulation of oceanic and atmospheric flows still requires approximations and simplifications of the full model. The barotropic vorticity equation (BVE) describing the single-layer QG equation with dissipative and forcing terms is one of the most commonly used models for the double-gyre wind-driven geophysical flows \cite{majda2006nonlinear}.

The BVE model shares many features with the two-dimensional Euler and Navier-Stokes equations and has been extensively used over the years to describe various aspects of the largest scales of turbulent geophysical fluid dynamics \cite{holland1980example, munk1982observing, griffa1989wind, cummins1992inertial,greatbatch2000four,nadiga2001dispersive}. Using $\beta-$plane assumption reasonable for most oceanic flows, the dimensionless vorticity-streamfunction formulation of the forced-dissipative BVE can be written as \cite{rahman2018hybrid}:
\begin{equation}\label{eq:nbve}
\frac{\partial \omega}{\partial t} + J(\omega,\psi) -\frac{1}{\mbox{Ro}}\frac{\partial \psi}{\partial x} = \frac{1}{\mbox{Re}}\nabla^2 \omega + \frac{1}{\mbox{Ro}}\sin(\pi y),
\end{equation}
where $\nabla^2$ is the standard two-dimensional Laplacian operator. $\omega$ and $\psi$ are the kinematic vorticity and streamfunction, respectively, defined as:
\begin{equation} \label{eq:omegadfn}
\omega = \nabla \times \mathbf{u},
\end{equation}
\begin{equation} \label{eq:psidfn}
\mathbf{u} = \nabla \times \psi \hat{k},
\end{equation}
where $\mathbf{u}$ is the two-dimensional velocity field and $\hat{k}$ refers to the unit vector perpendicular to the horizontal plane. The kinematic equation connecting the vorticity and streamfunction can be found by substituting the velocity components in terms of streamfunction in Equation~(\ref{eq:omegadfn}), which yields the following Poisson equation:
\begin{equation}\label{eq:pois}
\nabla^2 \psi = -\omega.
\end{equation}

Equation~(\ref{eq:nbve}) contains two dimensionless parameters, Reynolds number (Re) and Rossby number (Ro), which are related to the physical parameters and non-dimensional variables in the following way:
\begin{equation}\label{eq:ReRo1}
\mbox{Re} = \frac{V L}{\nu}, \quad \mbox{Ro} = \frac{V}{\beta L^2},
\end{equation}
where $\nu$ is the horizontal eddy viscosity of the BVE model and $\beta$ is the gradient of the Coriolis parameter at the basin center ($y = 0$). $L$ is the basin length scale and $V$ is the velocity scale, also known as the Sverdrup velocity \cite{sverdrup1947wind}, and is given by
\begin{equation}\label{eq:sverdrup}
V = \frac{\tau_0}{\rho H}\frac{\pi}{\beta L},
\end{equation}
where $\tau_0$ is the maximum amplitude of the double-gyre wind stress, $\rho$ is the mean fluid density, and $H$ is the mean depth of the ocean basin.
Despite not being explicitly represented in Equation~(\ref{eq:nbve}), there are two important relevant physical parameters, the Rhines scale, $\delta_I$, and the Munk scale, $\delta_M$, which are the non-dimensional boundary layer
thicknesses for the inertial and viscous (Munk) layer of the basin geometry, respectively. As a physical interpretation of these parameters in BVE model,  $\delta_I$ accounts for the strength of nonlinearity and $\delta_M$ is a measure of dissipation strength. $\delta_I$ and $\delta_M$ can be defined as
\begin{equation}
\dfrac{\delta_I}{L}=\left(\dfrac{V}{\beta L^2}\right)^{\frac{1}{2}} , \quad
\dfrac{\delta_M}{L}=\left(\dfrac{\nu}{\beta L^3}\right)^{\frac{1}{3}}
\end{equation}
and are related to Ro and Re by the following relations
\begin{equation}
\dfrac{\delta_I}{L}=\left(\mbox{Ro}\right)^{\frac{1}{2}} , \quad \dfrac{\delta_M}{L} = \left(\dfrac{\mbox{Ro}}{\mbox{Re}}\right)^{\frac{1}{3}}.
\end{equation}

Finally, the nonlinear advection term in Equation~(\ref{eq:nbve}) is given by the Jacobian
\begin{equation}\label{eq:jac}
J(\omega,\psi) =  \frac{\partial \psi}{\partial y}\frac{\partial \omega}{\partial x} - \frac{\partial \psi}{\partial x}\frac{\partial \omega}{\partial y}.
\end{equation}
Since ocean circulation models where the Munk and Rhines scales are close to each other, like the QG model, remain time dependent rather being converged to a steady state as time approaches to infinity \cite{medjo2000numerical}, numerical computations of these models are conducted in a statistically steady state, also known as the quasi-stationary state. Hence, in our study, we utilize numerical schemes suited for simulation of such type of ocean models and for long-time integration. Details of the relevant numerical discretization schemes, Poisson solver, and problem definitions for this study can be found in elsewhere~\cite{rahman2018hybrid,san2011approximate,san2016numerical}.

\section{Intrusive ROM-GP Methodology}
\label{sec:romgp}
The intrusive ROM framework is developed based on a standard Galerkin projection methodology using the method of snapshots, an efficient method for computing the POD basis functions \cite{sirovich1987turbulence}. In this section, we give a brief idea on the ROM-GP framework utilized in our work. We obtain $N$ number of snapshots for vorticity field, $\omega(x,y,t_n)$ for $n = 1,2,...,N$ at pseudo-time $t = t_n$ from full order model simulation (FOM). Algorithm~\ref{alg:pod} describes the POD basis construction procedure from the stored snapshots.

We can approximate the field variables, i.e., kinematic vorticity and streamfunction using the most energetic $R$ POD modes, where $R << N$, such that these $R$ largest energy containing modes correspond to the largest eigenvalues ($\lambda_1,...,\lambda_R$). The resulting full expression for the field variables can be written as:
\begin{align}
	\omega(x,y,t) = \bar{\omega}(x,y) + \sum_{k=1}^{R} a_k(t) \phi_k(x,y), \label{eq:fom_proj1} \\
    \psi(x,y,t) = \bar{\psi}(x,y) + \sum_{k=1}^{R} a_k(t) \theta_k(x,y), \label{eq:fom_proj2}
\end{align}
where $a_k(t)$ accounts for both streamfunction and vorticity based on the kinematic relation given by Eq.~(\ref{eq:pois}). It should be mentioned that in our ROM formulations, we use the following angle-parenthesis definition for the inner product of two arbitrary functions $f$ and $g$:
\begin{equation}\label{eq:inner}
    \int_{\Omega} f(x,y) g(x,y) dx dy = \langle f; g\rangle.
\end{equation}

\begin{algorithm}[H]
  \caption{POD basis construction}
  \label{alg:pod}
  \begin{algorithmic}[1]
    \State Compute the time-invariant mean fields and the fluctuation fields (mean-subtracted snapshots) for the given number of snapshots of the 2D vorticity field as:
    \Statex \begin{align}
        \bar{\omega}(x,y)&=\dfrac{1}{N}\sum_{n=1}^{N}\omega(x,y,t_n), \\
        \omega'(x,y,t_n)&=\omega(x,y,t_n)-\bar{\omega}(x,y).
    \end{align}
    \State An $N\times N$ snapshot data matrix $\mathbf{A}=[a_{ij}]$  is computed from the inner product of mean-subtracted snapshots
    \Statex \begin{equation}
        a_{ij}=\langle \omega'(x,y,t_i); \omega'(x,y,t_j)\rangle,
    \end{equation}
    \Statex where $i$ and $j$ refer to the snapshot indices.
    \State Compute the optimal POD basis functions by performing an eigendecomposition of $\mathbf{A}$ as $\mathbf{A} \mathbf{V} = \mathbf{V} \mathbf{\Lambda}$, where $\mathbf{\Lambda}$ is a diagonal matrix whose entries are the eigenvalues $\lambda_k$ of $\mathbf{A}$, and $\mathbf{V}$ is a matrix whose columns $\mathbf{v}_k$ are the corresponding eigenvectors. In our computations, we use the eigensystem solver based on the Jacobi transformations since $\mathbf{A}$ is a symmetric positive definite matrix \cite{press1992numerical}.
    \State Using the eigenvalues stored in a descending order (i.e., $\lambda_1\ge\lambda_2\ge\dots\ge\lambda_N$), for proper selection of the POD modes in $\boldsymbol\Lambda$, compute the orthogonal POD basis functions for the vorticity field $\phi_{k}$ as
    \Statex \begin{equation}\label{eq:podkv}
        \phi_{k}(x,y)=\dfrac{1}{\sqrt{\lambda_k}}\sum_{n=1}^{N} v^{n}_{k} \omega'(x,y,t_n),
    \end{equation}
    \Statex where $v^{n}_{k}$ is the $n^{th}$ component of the eigenvector $\mathbf{v}_k$. The scaling factor, $1/\sqrt{\lambda_k}$, is to guarantee the orthonormality of POD modes, i.e., $\langle \phi_i; \phi_j\rangle = \delta_{ij}$, where $\delta_{ij}$ is the Kronecker delta.
    \State Obtain the $k^{th}$ mode for the streamfunction, $\theta_k(x,y)$ utilizing the linear dependence between streamfunction and vorticity given by Equation~(\ref{eq:pois}):
    \Statex \begin{equation}\label{eq:podsf}
        \nabla^2 \theta_k = - \phi_k,
    \end{equation}
    \Statex for each $k=1, 2, ..., R$. To be able to use the same $a_{k}(t)$ coefficients for both streamfunction and vorticity fields, the following elliptic equation holds true for the mean variables:
       \Statex \begin{equation}\label{eq:podsf}
        \nabla^2 \bar{\psi} = - \bar{\omega}.
    \end{equation}
    \State Construct $k^{th}$ time-dependent modal coefficients $a_k(t_n)$ for $N$ snapshots by using POD modes and forward transformation:
    \Statex \begin{align}
	    a_k(t_n) = \langle\omega(x,y,t_n)-\bar{\omega}(x,y);\phi_k\rangle.
	\end{align}
  \end{algorithmic}
\end{algorithm}

We refer to \cite{rahman2019dynamic} for the details of the integration technique utilized in this study. In conventional projection based intrusive ROM framework, we apply Galerkin Projection to the governing equation, which yields $R$ coupled ordinary differential equations (ODEs) for the time evolution of the temporal modes of the system while the spatial modes are kept constant \cite{kunisch2002galerkin,rowley2004model,aubry1988dynamics}. Any standard time integration technique can be utilized to solve the coupled ODE system, since the basis functions and corresponding modal coefficients will be precomputed in the offline computation stage. The Galerkin projection approach is summarized in Algorithm~\ref{alg:gp}.

\begin{algorithm}[H]
  \caption{Galerkin projection to obtain ROM}
  \label{alg:gp}
  \begin{algorithmic}[1]
    \State Given an initial condition $\omega(x,y,t_0)$ at time $t_0$, compute the initial modal coefficients $a_k(t_0)$ using the relation below:
    \Statex \begin{align}
	    a_k(t_0) = \langle\omega(x,y,t_0)-\bar{\omega}(x,y);\phi_k\rangle.
	\end{align}
    \State Perform an orthogonal Galerkin projection by multiplying the governing equation with the POD basis functions and integrating over the domain \cite{rapun2010reduced}, which will yield the following dynamical system for $a_k$:
    \Statex \begin{equation}
        \frac{da_k}{dt} = \mathfrak{B}_{k} + \sum_{i=1}^{R} \mathfrak{L}^{i}_{k} a_i + \sum_{i=1}^{R}\sum_{j=1}^{R} \mathfrak{N}^{ij}_{k}a_i a_j,
	\end{equation}
    \Statex where $k=1,2, ..., R$ and the predetermined model coefficients can be computed by the following numerical integration (offline computing):
    \Statex \begin{eqnarray}
     \quad \mathfrak{B}_{k} &=& \big\langle  - J(\bar{\omega},\bar{\psi}) +\frac{1}{\mbox{Ro}}(\sin(\pi y) + \frac{\partial \bar{\psi}}{\partial x}) + \frac{1}{\mbox{Re}}\nabla^2 \bar{\omega}; \phi_{k} \big\rangle , \nonumber \\
      \quad \mathfrak{L}^{i}_{k} &=& \big\langle - J(\bar{\omega},\theta_{i}) -  J(\phi_{i},\bar{\psi}) +  \frac{1}{\mbox{Ro}}\frac{\partial \theta_{i}}{\partial x} + \frac{1}{\mbox{Re}}\nabla^2 \phi_{i}; \phi_{k} \big\rangle  , \nonumber\\
      \quad \mathfrak{N}^{ij}_{k} &=&  \big\langle -J(\phi_{i},\theta_{j}); \phi_{k} \big\rangle. \label{eq:roma7}
    \end{eqnarray}
  \end{algorithmic}
\end{algorithm}

\section{Non-intrusive ROM-LSTM Methodology}
\label{sec:romlstm}
In this section, we discuss our proposed ROM-LSTM methodology. As outlined in Algorithm~\ref{alg:pod}, we obtain the time-dependent modal coefficients $a_k$ by performing a POD transform on stored snapshot data. The modal coefficients are a sequence of data points with respect to time, i.e., a time series representing the underlying dynamical system. In intrusive or physics-based ROM, we do Galerkin projection using governing equation to obtain a coupled system of ODEs for $a_k$, and then solve the ODE system on the given time interval. However, the limitations of projection based ROMs, such as susceptibility to instability for noisy data set, numerical constraints for solving ODE system, or inefficient reduced order modeling, encourage us to replace the physics-based Galerkin projection phase of ROM-GP methodology with a data-driven approach. Among the variety of ideas to resolve the issues associated with projection based ROM, a number of published works related to ROM based on POD and neural networks have shown signs of future success. The recurrent neural network (RNN) is one of the widely used neural network architectures in ROMs which is designed to operate on input information as well as the previously stored observations to predict the dependencies among the temporal data sequences \cite{jaeger2004harnessing,lecun2015deep}. LSTM is a special variant of RNN which is capable of tracking long-term dependencies among the input data sequences. Hence, we consider LSTM recurrent neural network to develop our non-intrusive ROM-LSTM framework. Before describing the ROM-LSTM procedure, we first briefly review the LSTM architecture.

As the name suggests, RNNs contain recurrent or cyclic connections that enable them to model complex time-varying data sequences with a wide range of temporal dependencies or correlations between them. In general, RNNs map a sequence of data to another sequence through time using cyclic connections, and constrain some of the connections to hold the same weights using back-propagation algorithm \cite{rumelhart1988learning}. However, the standard RNN architecture suffers from vanishing gradient problem when the gradient of some weights starts to become too small or too large \cite{bengio1994learning}. This leads to the development of improved RNN architectures which overcome the modeling issues of standard RNNs. One of the most successful forms of improved RNN architectures is the LSTM network, which solves the limitation of vanishing gradients \cite{hochreiter1997long}. In contrast to most of the ANN architectures, LSTM operates by cell states and gating mechanisms to actively control the dynamics of cyclic connections and thus, resolves the vanishing gradient issues. Similar to the standard RNNs, LSTM can learn and predict the temporal dependencies based on the input information and previously acquired information, i.e., the internal memory of LSTM allows the network to find the relationship between the current input and stored information to make a prediction.

The conventional LSTM architecture contains memory blocks in the recurrent hidden layers, which have memory cells to store the cell states and gates to control the flow of information. Each memory block has an input gate controlling the flow of input activations into the cell, a forget gate to adaptively forgetting and resetting the cell's memory (to prevent over-fitting by processing continuous inflow of input streams), and the output gate controlling the output flow of cell activations into the next cell. In our LSTM architecture, we consider an input sequential data matrix $\mathcal{X}_k = \left\{a_k^{(1)}, a_k^{(2)}, ..., a_k^{(N-\sigma)}\right\}$ for training, where $\sigma$ is the lookback time-window. The lookback time-window, in our definition, means the time history over which the LSTM model does the training and prediction recursively. Indeed, increasing the value of $\sigma$ increases the quality of training the model, but makes the model dependent on an increased number of initial states during prediction. The LSTM model is trained to an output sequential data matrix $\mathcal{Y}_k = \left\{a_k^{(\sigma+1)}, a_k^{(\sigma+2)}, ..., a_k^{(N)}\right\}$ recursively (here, $k=1,2, ..., R$ and $\mathcal{X}, \mathcal{Y} \in a$). Considering input gate as $\mathcal{I}$, the forget gate as $\mathcal{F}$, the output gate as $\mathcal{O}$, the cell activation vectors as $c$, and the cell output as $h$, the LSTM model does the mapping from the input states to an output state by using the following equations at any instance $n$ \cite{sak2014long,yeo2019deep}:
\begin{align}\label{eq:lstm}
	\text{Input network:} \quad &\mathcal{X}_k^{(n)} =  \text{tanh}\left(W_h h_k^{(n)} + W_{\mathcal{Y}} \mathcal{Y}_k^{(n)}\right), \\
    \text{Gate functions:} \quad &m_k^{(n)} =  \zeta\left(W_m \mathcal{X}_k^{(n)} + b_m\right), \\
                                  &\text{for} \ \ m \in (\mathcal{I}, \mathcal{F}, \mathcal{O}) \nonumber \\
    \text{Internal cell state:} \quad &c_k^{(n+1)} =  \mathcal{F}_k^{(n)}\odot c_k^{(n)} + \mathcal{I}_k^{(n)} \odot \xi, \\
                                     &\text{where} \ \ \xi =  \text{tanh}\left(W_c \mathcal{X}_k^{(n)} + b_c\right) \nonumber \\
    \text{Output state:} \quad &h_k^{(n+1)} =  \mathcal{O}_k^{(n)} \odot \text{tanh}\left(c_k^{(n+1)}\right), \\
    \text{Output network:} \quad &\widehat{\mathcal{Y}_k}^{(n+1)} =  W_{\mathcal{Y}2}\text{tanh}\big(W_{\mathcal{Y}1} h_k^{(n+1)}  \nonumber  \\
                                    & \quad \quad \quad \quad + b_{\mathcal{Y}1}\big) + b_{\mathcal{Y}2},
\end{align}
where $W$ represents the weight matrices for each gates, $\widehat{\mathcal{Y}_k}^{(n+1)}$ is the LSTM prediction, $b$ denotes the bias vectors for each gates, $\odot$ is the element-wise product of vectors, and $\zeta$ is the logistic sigmoid function.

\begin{figure*}[htbp]
\centering
\includegraphics[width=\mywd]{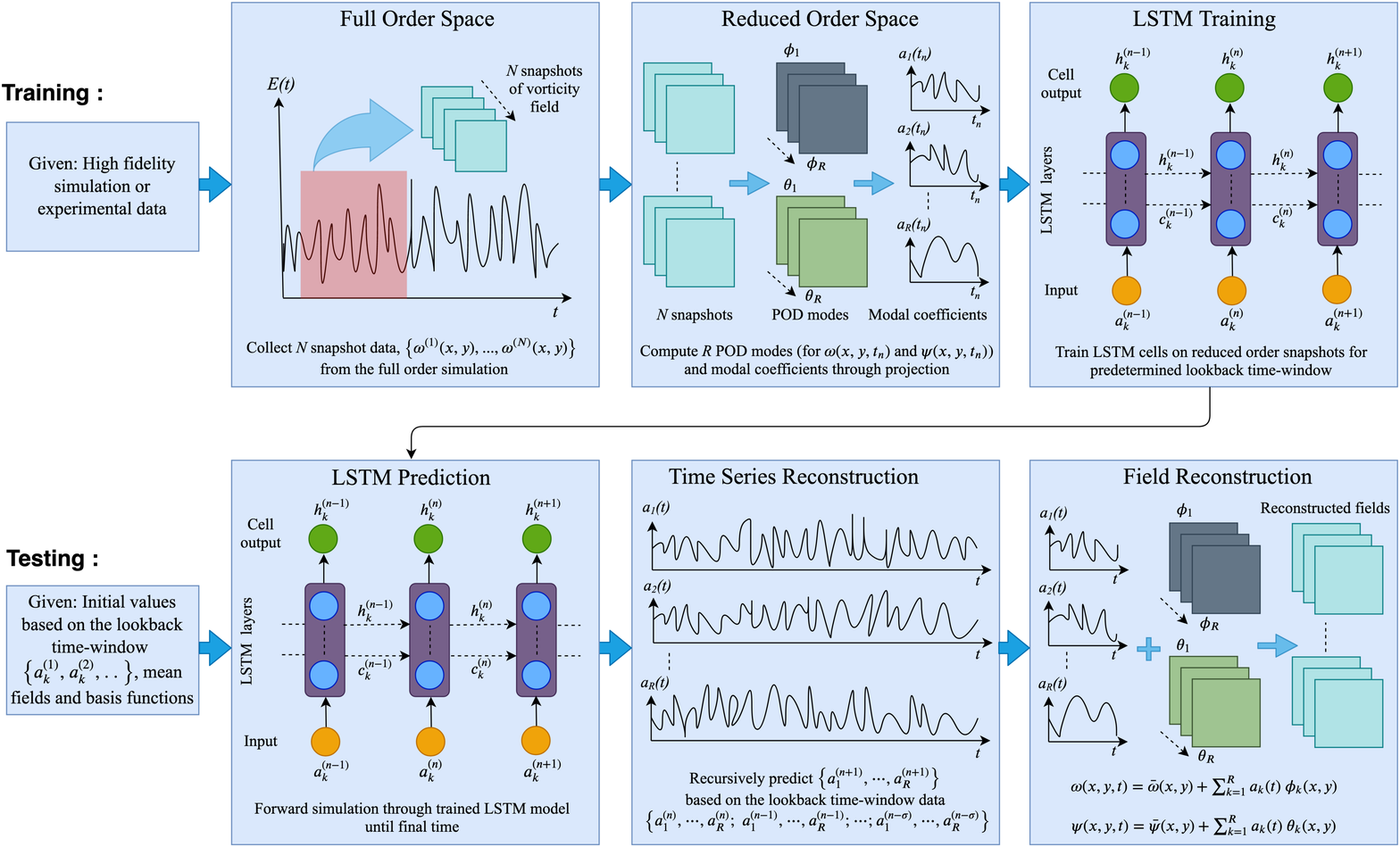}
\caption{Workflow diagram of the ROM-LSTM framework. Note that the training phase (offline computation) is computationally heavier compared to the testing (online computation) phase.}
\label{fig:1}
\end{figure*}

Similar to the ROM-GP methodology, the workflow of the ROM-LSTM framework consists of two phases as displayed in Figure~\ref{fig:1}. In the offline training phase, we first obtain POD basis functions and modal coefficients using Algorithm~\ref{alg:pod}. The known time series of modal coefficients from training snapshots are used to train the LSTM model. Based on the values of $\sigma$, the input of the LSTM model $\mathcal{M}$ will be the previous time states of the input modal coefficients for $R$ retained modes and the output of the model will be the next time state recursively for $R$ modes. Training LSTM model is the computationally heavier part of the ROM-LSTM framework, but this is done offline. In online testing phase, we recursively predict the modal coefficients for the total time using the trained model $\mathcal{M}$. The input of the trained model $\mathcal{M}$ will be the initial states $\left\{a_{1}^{(1)} ,\dots, a_{R}^{(1)}; \dots; a_{1}^{(\sigma)} ,\dots, a_{R}^{(\sigma)}\right\}$ based on the preselected value of $\sigma$ and the output will be recursive prediction of corresponding future time states. Thus, we bypass the physics-based Galerkin projection part with completely data-driven neural network approach to predict the modal coefficients. Also, the computational cost of prediction through trained LSTM network is significantly lower than the physics-based approach. Finally, we reconstruct the mean vorticity and streamfunction fields using inverse transform to analyze the behavior of the quasi-stationary flow. The key steps of the ROM-LSTM framework are outlined below in Algorithm~\ref{alg:lstm}.

To design our LSTM architecture for ROM-LSTM framework, we utilize Keras \cite{chollet2015keras}, a high level API designed for deep learning, combined with standard Python libraries. The FOM simulation for data snapshots generation and POD basis construction codes are written in FORTRAN programming language. We utilize a deep LSTM network stacking 6 LSTM layers with 40 neurons in each layer. The computational cost is found manageable in this deep architecture setup, which encourages us to perform all the numerical experiments with this same setup. The mean-squared error (MSE) is chosen as the loss function, and a variant of stochastic gradient descent method, called ADAM \cite{kingma2014adam}, is used to optimize the mean-squared loss. The other relevant hyper-parameters utilized in our LSTM architecture are documented in Table.~\ref{tab:1}. The hyper-parameters are kept constant for all simulations to obtain a fair comparison between the results in different numerical experiment runs. It should be noted that the training data is normalized by the minimum and maximum of each time series to be in between the range $[-1, +1]$.

\begin{algorithm}[H]
  \caption{ROM-LSTM framework}
  \label{alg:lstm}
  \begin{algorithmic}[1]
    \Statex \underline{\textit{Training (offline)}} \smallskip
        \State Collect $N$ snapshot data for the vorticity field, $\omega(x,y,t_n) = \left\{\omega^{(1)}(x,y), \omega^{(2)}(x,y), ... ,\omega^{(N)}(x,y)\right\}$ from the FOM simulation.
        \State Compute $R$ POD modes for kinematic vorticity, $\phi_k$  and streamfunction, $\theta_k$ using Equation~(\ref{eq:podkv}) and Equation~(\ref{eq:podsf}), respectively, for $k=1,2, ..., R$.
        \State Construct modal coefficients by a forward transform through projection
        \Statex \begin{equation}
            a_k(t_n) = \langle\omega(x,y,t_n)-\bar{\omega}(x,y);\phi_k\rangle,
        \end{equation}
        \Statex where $a_k(t_n) = \left\{a^{(1)}_k, a^{(2)}_k ,\dots, a^{(N)}_k\right\}$.
        \State Train LSTM model on reduced order snapshots for selected lookback time-window $\sigma$:
        \Statex \begin{align}
           \quad &\mathcal{M}: \left\{a_{1}^{(n)} ,\dots, a_{R}^{(n)}; \ \dots \ ; a_{1}^{(n-\sigma)},\dots, a_{R}^{(n-\sigma)}\right\} \nonumber \\
           \quad &\Rightarrow \left\{a_{1}^{(n+1)} ,\dots, a_{R}^{(n+1)}\right\}.
        \end{align}

    \Statex \underline{\textit{Testing (Online)}} \smallskip
        \State Given initial values $\{a_k^{(1)}, a_k^{(2)} ,\dots, a_{k}^{(\sigma)}\}$ based on $\sigma$, precomputed mean values and basis functions.
        \State Use the trained LSTM model $\mathcal{M}$ to recursively predict $a_k(t)$ until final time reached.
        \State Reconstruct the mean fields by inverse transform using Equation~(\ref{eq:fom_proj1}) and Equation~(\ref{eq:fom_proj2}).
  \end{algorithmic}
\end{algorithm}

\begin{table}[htbp]
\centering
\caption{A list of hyper-parameters utilized to train the LSTM network for all numerical experiments.}
\begin{tabular}{p{0.32\textwidth}p{0.16\textwidth}}
\hline\noalign{\smallskip}
Parameters & \quad \quad \quad Values \\
\noalign{\smallskip}\hline\noalign{\smallskip}
Number of hidden layers  & \quad \quad \quad 6  \\
Number of neurons in each hidden layer  & \quad \quad \quad 40  \\
Batch size & \quad \quad \quad 16 \\
Epochs & \quad \quad \quad 500 \\
Activation functions in the LSTM layers & \quad \quad \quad tanh \\
Training-testing ratio & \quad \quad \quad 4:9 \\
Validation data set & \quad \quad \quad 20$\%$ \\
Loss function &  \quad \quad \quad MSE \\
Optimizer & \quad \quad \quad ADAM \\
\noalign{\smallskip}\hline
\end{tabular}
\label{tab:1}
\end{table}

\section{Numerical Results}
\label{sec:results}
The predictive performance of the ROM-LSTM framework is thoroughly examined in this section in terms of time series evolution of the modal coefficients and mean flow fields. It is well documented in literature that the ROM-GP framework is incapable of capturing mean flow dynamics for quasi-stationary flows using lower number of POD modes, and susceptible to instability \cite{san2014proper,san2015stabilized}. There have been a number of approaches proposed in previous literature to improve the ROM performance. One way to stabilize the ROM is by adding an empirical stabilization parameter based on the analogy between large eddy simulation and truncated modal projection \cite{aubry1988dynamics,wang2012proper}. Later, it is found that the ROM performance further improves taking the optimal value for the stabilization parameter rather than selecting it arbitrarily \cite{rempfer1991,san2014proper,cazemier1997,san2015stabilized}. In our previous work, we have shown that computing the stabilization parameter dynamically at each time step improve the ROM performance significantly \cite{rahman2019dynamic}. 
However, the proposed ROM-LSTM methodology has several advantages over the physics-based approaches, such as, no dependence on the underlying governing dynamical system to obtain the solution, i.e., the process is free of numerical constraints, no burden of adding stabilization parameter to account for instability issues and so on.
To reach a conclusion about the performance of the ROM-LSTM framework, we compare ROM-LSTM predictions with the FOM simulation and the standard ROM-GP results. Moreover, we present the performance of the ROMs based on lookback time-window $\sigma$ and LSTM training for different number of POD modes to show the robustness and capability of the proposed framework. Furthermore, we present the L\textsubscript{2}-norm errors to perform a quantitative assessment on the accuracy of the ROM-LSTM solutions with respect to ROM-GP solutions.

We choose the single-layer QG problem as our test bed to evaluate the performance of ROMs. Because of the complex flow behavior with wide range of scales, QG problem has been utilized as test problem in many notable literature \cite{cushman2011introduction,holm2003modeling,san2011approximate,cummins1992inertial,greatbatch2000four}. To make the analyses simple and easily understandable, we present simulation results only for Re$=450$ and Ro$=3.6 \times 10^{-3}$ flow condition, which can be considered turbulent enough and suitable for reduced order modeling. The FOM simulation is done from $t = 0$ to $t=100$ using a constant time step of $\Delta t = 2.5\times10^{-4}$ on a Munk layer resolving $256\times512$ grid resolution (i.e., consisting of about four grid points in the Munk scale, i.e., $\delta_M/L = 0.02$). To avoid the initial transient time interval, we store $400$ data snapshots from $t = 10$ to $t=50$ to generate the POD bases and modal coefficients to train the LSTM model. We refer to Ref.~\cite{rahman2019dynamic} to get an idea on the POD analysis as well as the instantaneous vorticity field plots for the same flow condition. The computational domain of our test problem is $(x,y) \in [0, 1]\times[-1, 1]$. To understand the nature of the QG data set, we compute the Hurst exponent, $H$, for the modal coefficients. The Hurst exponent is a statistical measure of the presence of long-term trends in a non-stationary time series \cite{hurst1951long}. Thus, the Hurst exponent can help in selecting the appropriate model for a given time series prediction. We also note that the Hurst exponent has been utilized in many research fields, e.g., hydrology, finance, and healthcare indsutry \cite{chandra2013telecardiology,koutsoyiannis2003climate,lahmiri2014new,bassler2006markov,carbone2004time}. $H$ can be statistically defined as \cite{qian2004hurst}:
\begin{equation} \label{hurst}
\text{E} \left[ \frac{\text{Range}(n)}{\text{SD} (n)}\right] = k n^H, \quad \text{as} \ n \rightarrow \infty.
\end{equation}
Here, E is the expected value of the ratio between the range of the first $n$ cumulative deviations from the mean and their corresponding standard deviations (SD), $n$ is the time span of the observations, and $k$ is constant. The range of $H$ is in between 0 and 1. $H \rightarrow 1$ means a persistent series (a strong trend in the time series at hand), $H \rightarrow 0$ means an anti-persistent series (a time series with long-term switching between high and low values) and $H \approx 0.5$ indicates a random series (fewer correlations between current and future observations). Interested readers are directed to Ref.~\cite{mohan2018deep} for a detailed description of suitability of LSTM as a predictive modeling approach for different time series data using the measurement of $H$. We calculate the $H$ for modal coefficients of QG data set for given flow conditions using the so called rescaled range (R/S) analysis, popularized by \citeauthor{mandelbrot1968noah} \cite{mandelbrot1968noah,mandelbrot1969robustness}. The details of (R/S) analysis can be found in Ref.~\cite{qian2004hurst}. The Hurst exponents for the modal coefficients of QG case are tabulated in Table~\ref{tab:2}, where we can see that the values of $H$ are around 0.5. This indicates the randomness of the QG problem, which can be a good representative of large-scale quasi-stationary  geophysical turbulent flow systems.

\begin{table}[htbp]
\centering
\caption{Hurst exponents of modal coefficients.}
\begin{tabular}{p{0.20\textwidth}p{0.19\textwidth}}\\
\hline\noalign{\smallskip}
Modal coefficient & Hurst exponent\\\hline\noalign{\smallskip}
\quad $a_1(t)$ &  \quad 0.52 \\
\quad $a_2(t)$ &  \quad 0.35 \\
\quad $a_3(t)$ &  \quad 0.63 \\
\quad $a_4(t)$ &  \quad 0.59 \\
\quad $a_5(t)$ &  \quad 0.49 \\
\quad $a_6(t)$ &  \quad 0.59 \\
\quad $a_7(t)$ &  \quad 0.59 \\
\quad $a_8(t)$ &  \quad 0.46 \\
\quad $a_9(t)$ &  \quad 0.58 \\
\quad $a_{10}(t)$ &  \quad 0.53 \\
\hline
\end{tabular}
\label{tab:2}
\end{table}

\begin{figure*}[htbp]
\centering
\includegraphics[width=\mywd]{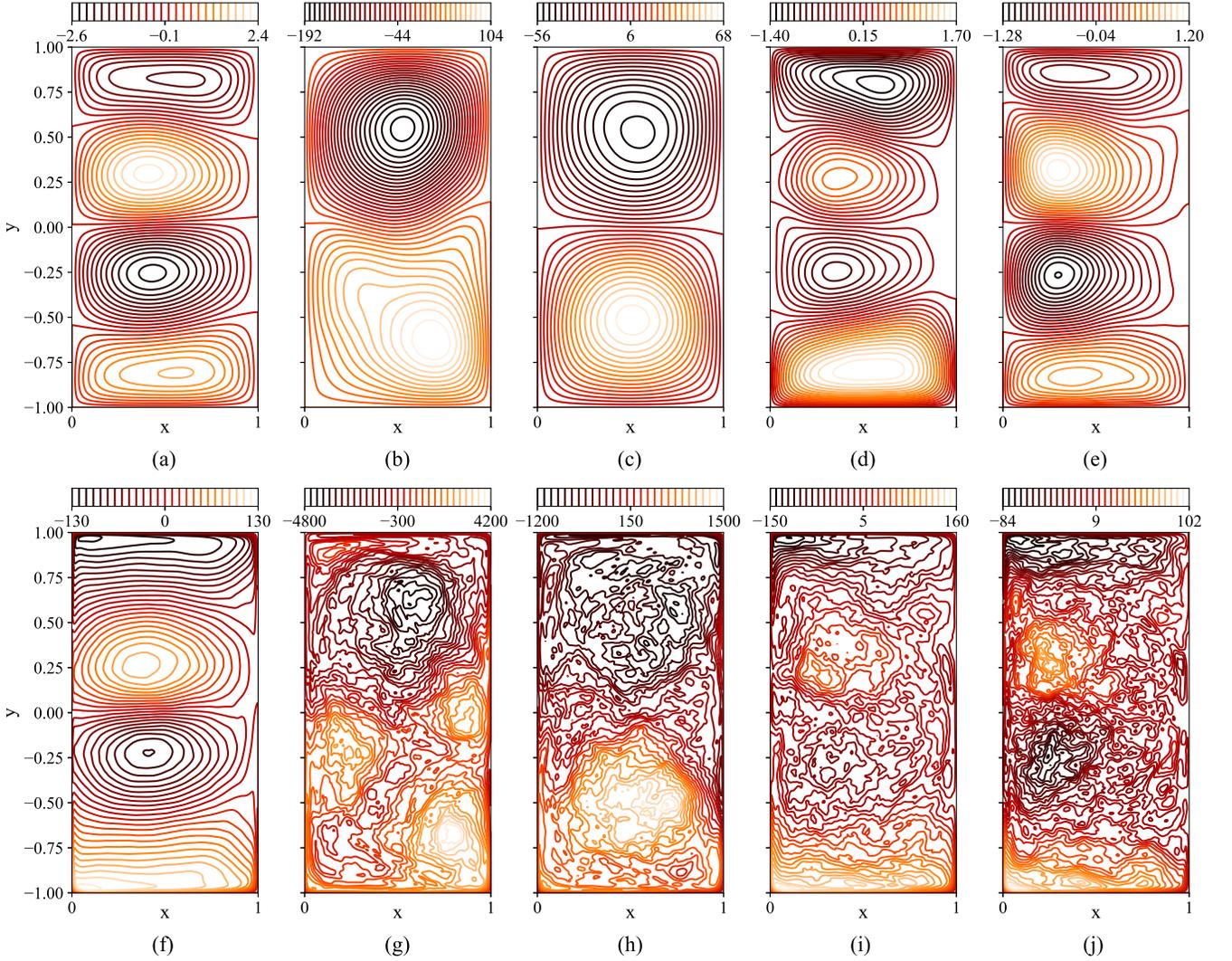}
\caption{Mean streamfunction and vorticity fields obtained by the FOM simulation and the standard ROM-GP simulation at Re $=450$ and Ro $=3.6 \times 10^{-3}$ flow condition. (a) $\psi_{\text{FOM}}$ at a resolution of $256 \times 512$, (b) $\psi_{\text{ROM-GP}}$ with $R = 10$ modes, (c) $\psi_{\text{ROM-GP}}$ with $R = 20$ modes, (d) $\psi_{\text{ROM-GP}}$ with $R = 40$ modes, (e) $\psi_{\text{ROM-GP}}$ with $R = 80$ modes, (f) $\omega_{\text{FOM}}$ at a resolution of $256 \times 512$, (g) $\omega_{\text{ROM-GP}}$ with $R = 10$ modes, (h) $\omega_{\text{ROM-GP}}$ with $R = 20$ modes, (i) $\omega_{\text{ROM-GP}}$ with $R = 40$ modes, (j) $\omega_{\text{ROM-GP}}$ with $R = 80$ modes.}
\label{fig:2}
\end{figure*}

\begin{figure*}[htbp]
\centering
\includegraphics[width=\mywd]{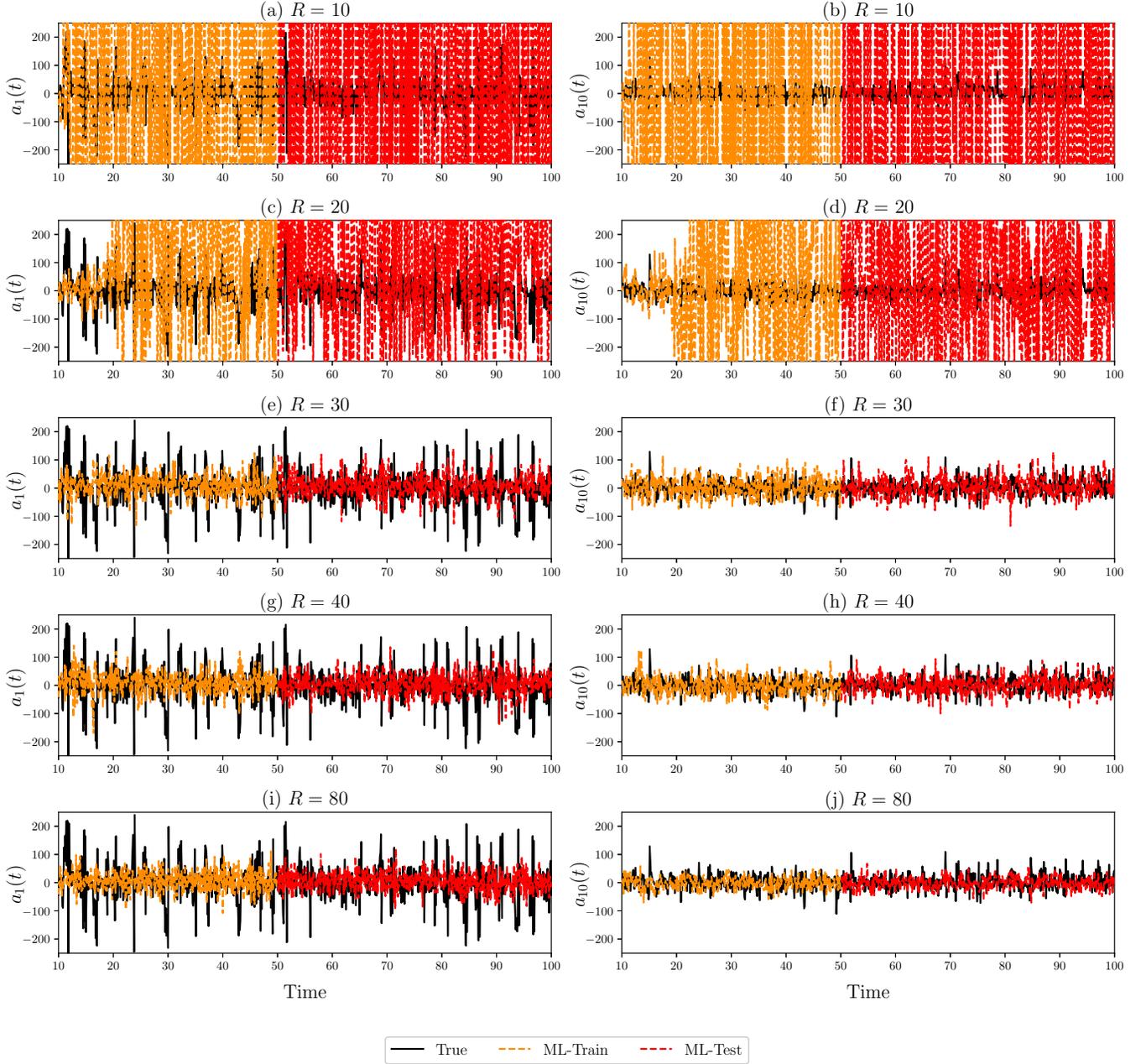}
\caption{Time series evolution of the first and tenth modal coefficients, $a_1 (t)$ and  $a_{10} (t)$ respectively, between $t = 10$ to $t = 100$ for standard ROM-GP simulation at Re $=450$ and Ro $=3.6 \times 10^{-3}$. (a) $a_1 (t)$ for ROM-GP with $R = 10$ modes, (b) $a_{10} (t)$ for ROM-GP with $R = 10$ modes, (c) $a_1 (t)$ for ROM-GP with $R = 20$ modes, (d) $a_{10} (t)$ for ROM-GP with $R = 20$ modes, (e) $a_1 (t)$ for ROM-GP with $R = 30$ modes, (f) $a_{10} (t)$ for ROM-GP with $R = 30$ modes, (g) $a_1 (t)$ for ROM-GP with $R = 40$ modes, (h) $a_{10} (t)$ for ROM-GP with $R = 40$ modes, (i) $a_1 (t)$ for ROM-GP with $R = 80$ modes, (j) $a_{10} (t)$ for ROM-GP with $R = 80$ modes. True projection series is underlined in each figure with black straight line. The training zone is shown with orange dashed line (from $t = 10$ to $t = 50$) and the out-of-sample testing zone is shown with red dashed line (from $t = 51$ to $t = 100$) in ROM-LSTM solution series in each figure.}
\label{fig:3}
\end{figure*}

Figure~\ref{fig:2} shows the mean streamfunction and vorticity field contours obtained by the ROM-GP model. To compare the predictive performance of the ROM-GP model with respect to the true solution, we include the mean contour plots of FOM simulation on the left column as well. We can see the full order solution displays a four-gyre circulation patterns for both mean streamfunction and vorticity fields. Since the instantaneous fields for the QG flow is always fluctuating in time, it becomes difficult to compare solutions of different models at the same time state. However, the mean fields always exhibit the four-gyre circulation for higher Re (highly turbulent regime, i.e., turbulence with weak dissipation) which implies a state of turbulent equilibrium between two inner gyres circulation representing the wind stress curl forcing and the outer gyres representing the eddy flux of potential vorticity (the northern and southern gyres found in geostrophic turbulence) \cite{greatbatch2000four}. In our study, the time-averaged (mean) field data are obtained by averaging between $t = 10$ to $t=100$. Another point to be noted in FOM field plots that the bright orange circulations in the four-gyres (top circulation of the inner gyres and bottom circulation of the outer gyres) indicate the circulation in counter clockwise or positive direction and the other two circulations signifies the circulation in clockwise direction. We can observe in Figure~\ref{fig:2} that the ROM-GP simulations with $R = 10$ and $R = 20$ modes display a non-physical two-gyre circulation for streamfunction whereas the vorticity field does not capture almost any conclusive physical pattern. However, the results improve with increasing modes as we can see the streamfunction contour is showing clear four-gyre patterns even though the vorticity plots are very noisy compared to the true solution. These observations are supported by the time series evolution of first and tenth modal coefficient plots in Figure~\ref{fig:3}. It is apparent that the increasing modes stabilize the noise to reach a physical solution for both modal coefficients.

We note that the time scale in our formulation is normalized by $L/V$ to obtain dimensionless time unit. Following \cite{san2013efficient}, typical oceanic values (e.g., $L=2000$ km and $\beta = 1.75 \times 10^{-11}$ m$^{-1}$s$^{-1}$)  yield approximately $L/V=0.25$ year for $\mbox{Ro}=0.0036$. Therefore, a numerical simulation over 100 computational time units refers to the evolution of flow dynamics over 25 years in physical time. Therefore, the intermittent bursts appeared in the true projection of the most energetic mode (i.e., $a_{1}(t)$ indicate the seasonal variations in QG dynamics. Although ROM-GP yields non-physical solution for $R=10$ and $R=20$ cases, $a_{k}(t)$ series reaches more meaningful levels for $R=30$ and beyond. However, it is hard to claim from Figure~\ref{fig:3} that the ROM-GP yields an accurate prediction of these seasonal bursts even for higher $R$ values.

\begin{figure*}[htbp]
\centering
\includegraphics[width=\mywd]{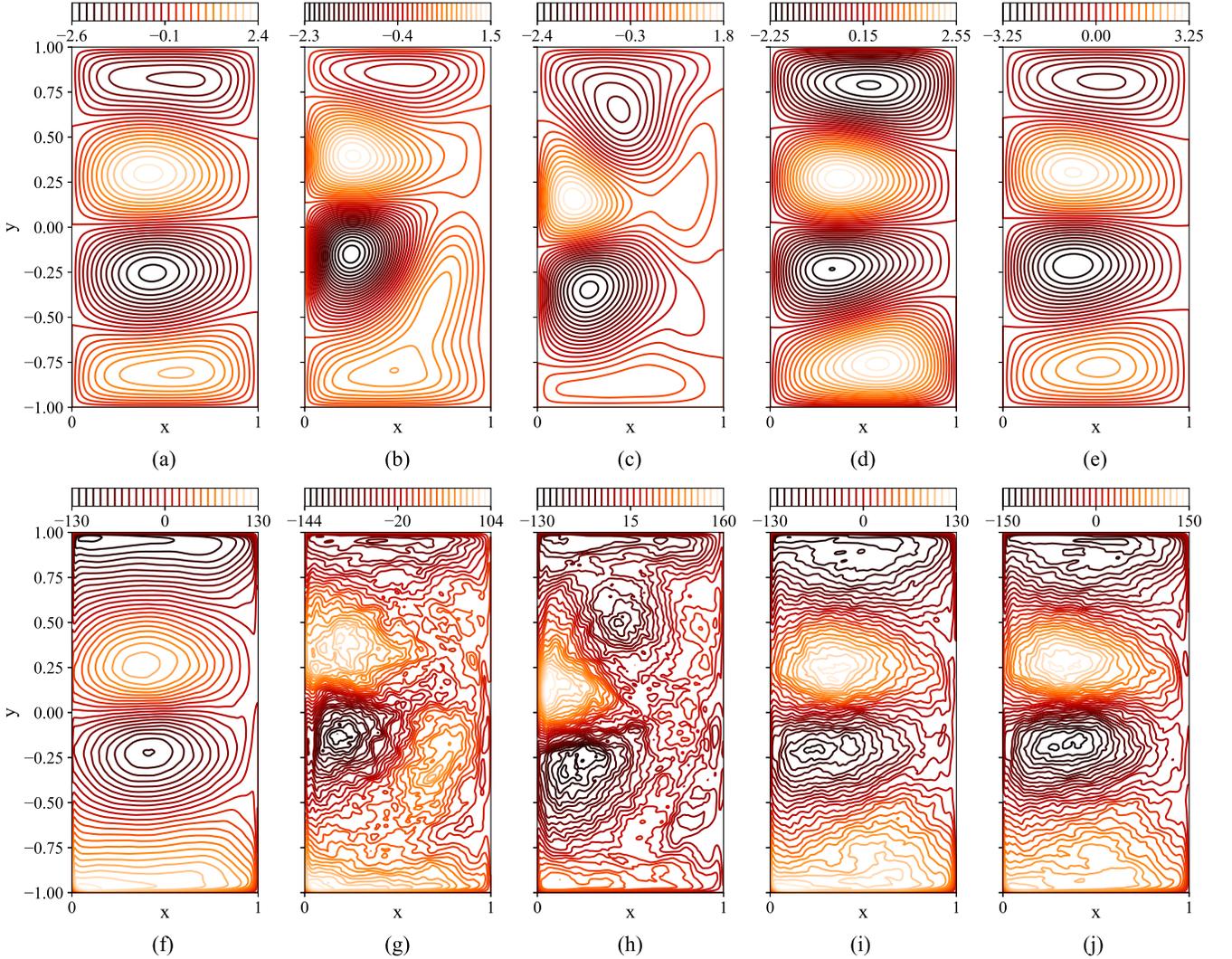}
\caption{Mean streamfunction and vorticity fields obtained by the ROM-LSTM simulation based on different lookback time-window, $\sigma$ at Re $=450$ and Ro $=3.6 \times 10^{-3}$ flow condition. (a) $\psi_{\text{FOM}}$ at a resolution of $256 \times 512$, (b) $\psi_{\text{ROM-LSTM}}$ with $\sigma= 1$, (c) $\psi_{\text{ROM-LSTM}}$ with $\sigma= 2$, (d) $\psi_{\text{ROM-LSTM}}$ with $\sigma= 4$, (e) $\psi_{\text{ROM-LSTM}}$ with $\sigma= 5$, (f) $\omega_{\text{FOM}}$ at a resolution of $256 \times 512$, (g) $\omega_{\text{ROM-LSTM}}$ with $\sigma= 1$, (h) $\omega_{\text{ROM-LSTM}}$ with $\sigma= 2$, (i) $\omega_{\text{ROM-LSTM}}$ with $\sigma= 4$, (j) $\omega_{\text{ROM-LSTM}}$ with $\sigma= 5$. Note that the LSTM model is trained with $R = 10$ modes.}
\label{fig:4}
\end{figure*}

\begin{figure*}[htbp]
\centering
\includegraphics[width=\mywd]{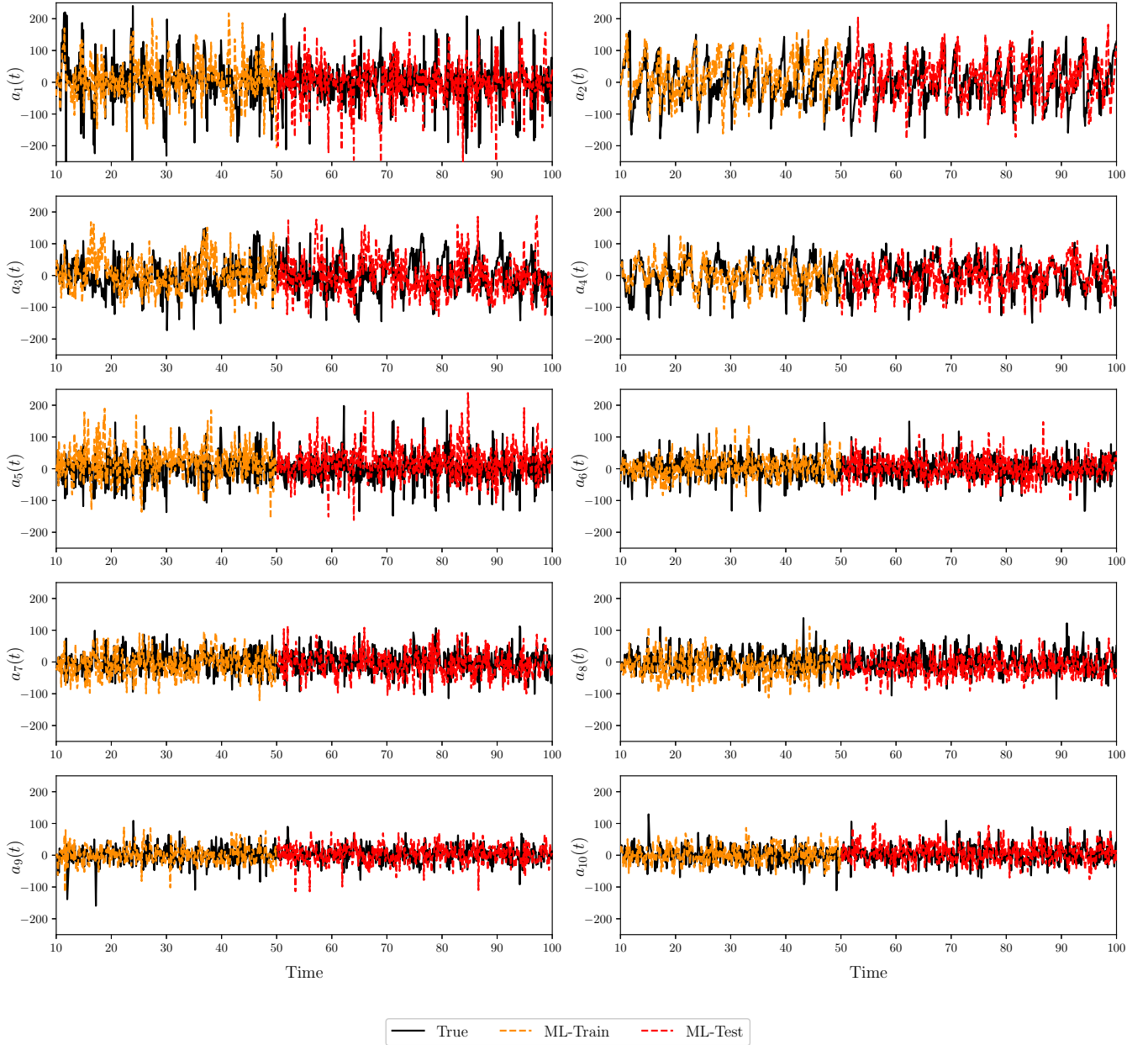}
\caption{Time series evolution of the modal coefficients between $t = 10$ to $t = 100$ for ROM-LSTM simulation at Re $=450$ and Ro $=3.6 \times 10^{-3}$. Note that the LSTM model is trained with $R = 10$ modes and $\sigma= 5$. True projection series is underlined in each figure with black straight line. The training zone is shown with orange dashed line (from $t = 10$ to $t = 50$) and the out-of-sample testing zone is shown with red dashed line (from $t = 51$ to $t = 100$) in ROM-LSTM solution series in each figure.}
\label{fig:5}
\end{figure*}

We present the field contours obtained by ROM-LSTM based on different $\sigma$ values in Figure~\ref{fig:4}. It can be seen that $\sigma = 1$ and $\sigma = 2$ do not provide much accurate results as the patterns get distorted in some extent even though they are being able to capture the four-gyre. However, both streamfunction and vorticity contours show a stable and accurate prediction of the true mean fields for $\sigma = 4$ and $\sigma = 5$. Though the vorticity field contour is not displaying as smooth contour lines as the true solution due to the reduction of dimension order, it is showing a better performance compared to the ROM-GP solutions. As shown in the recent work of \citeauthor{yeo2019short} \cite{yeo2019short}, the LSTM network trained on a noisy data learns to reduce the contributions of noisy input data by developing its own dynamics and thus, the prediction remains close to the truth rather than being unstable due to noisy input data. Hence, the LSTM prediction is expected to yield a stable and physical solution for a fluctuating quasi-stationary system. It should be noted that these results are obtained for LSTM training with $R = 10$ modes. The time series evolution plots for the modal coefficients based on $\sigma = 5$ and $R = 10$ modes in Figure~\ref{fig:5} show that ROM-LSTM time series predictions are almost on top of the true projection of modal coefficients. Even though the model is trained for $t = 10$ to $t=50$ only, the ROM-LSTM model is able to obtain a stable and accurate prediction up to the final time $t=100$.

\begin{figure*}[htbp]
\centering
\includegraphics[width=\mywd]{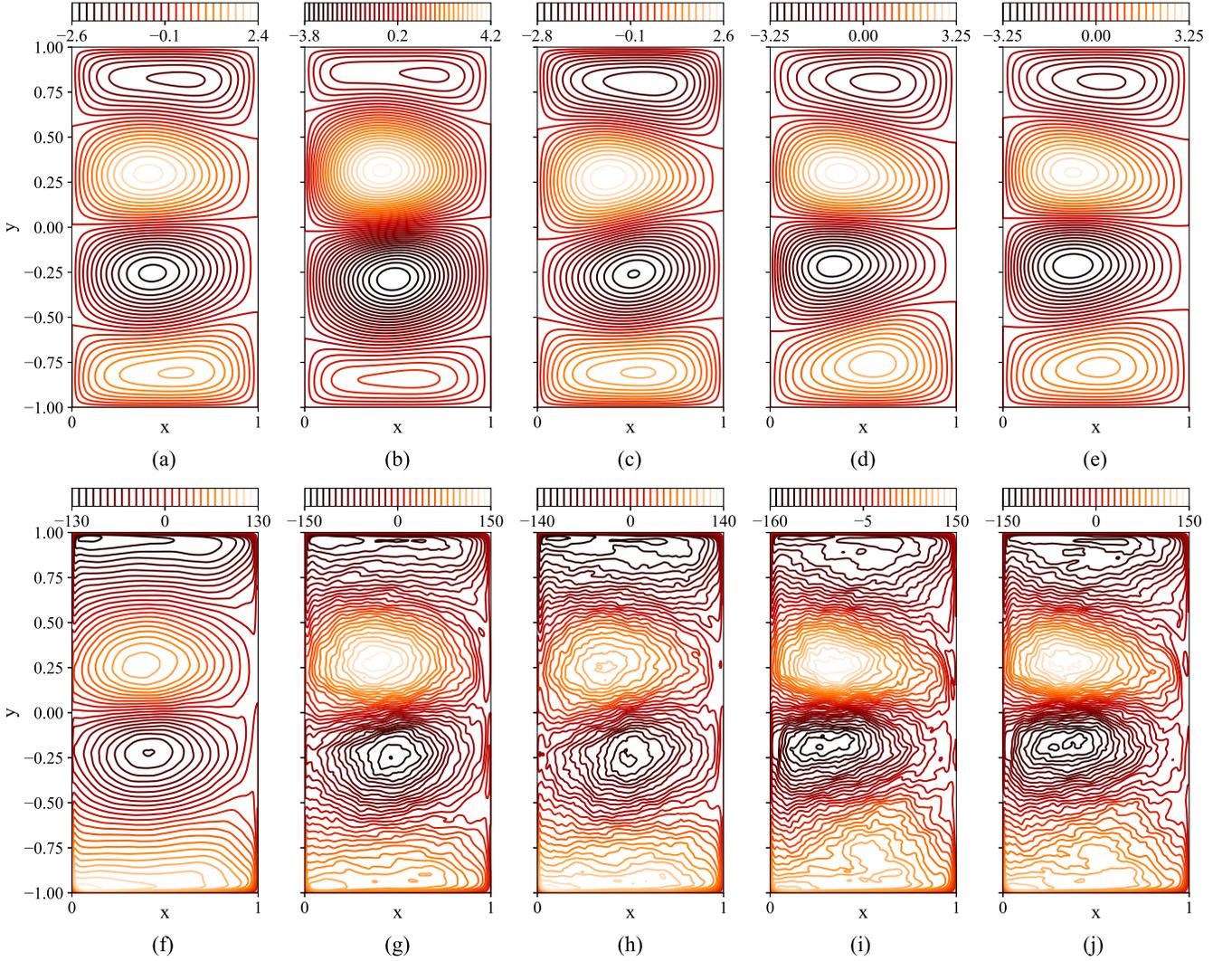}
\caption{Mean streamfunction and vorticity fields obtained by the ROM-LSTM simulation based on the number of modes to train the LSTM model at Re $=450$ and Ro $=3.6 \times 10^{-3}$ flow condition. (a) $\psi_{\text{FOM}}$ at a resolution of $256 \times 512$, (b) $\psi_{\text{ROM-LSTM}}$ for LSTM training with $R = 2$ modes, (c) $\psi_{\text{ROM-LSTM}}$ for LSTM training with $R = 4$ modes, (d) $\psi_{\text{ROM-LSTM}}$ for LSTM training with $R = 8$ modes, (e) $\psi_{\text{ROM-LSTM}}$ for LSTM training with $R = 10$ modes, (f) $\omega_{\text{FOM}}$ at a resolution of $256 \times 512$, (g) $\omega_{\text{ROM-LSTM}}$ for LSTM training with $R = 2$ modes, (h) $\omega_{\text{ROM-LSTM}}$ for LSTM training with $R = 4$ modes, (i) $\omega_{\text{ROM-LSTM}}$ for LSTM training with $R = 8$ modes, (j) $\omega_{\text{ROM-LSTM}}$ for LSTM training with $R = 10$ modes. Note that the LSTM model is trained with $\sigma= 5$.}
\label{fig:6}
\end{figure*}

\begin{figure*}[htbp]
\centering
\includegraphics[width=\mywd]{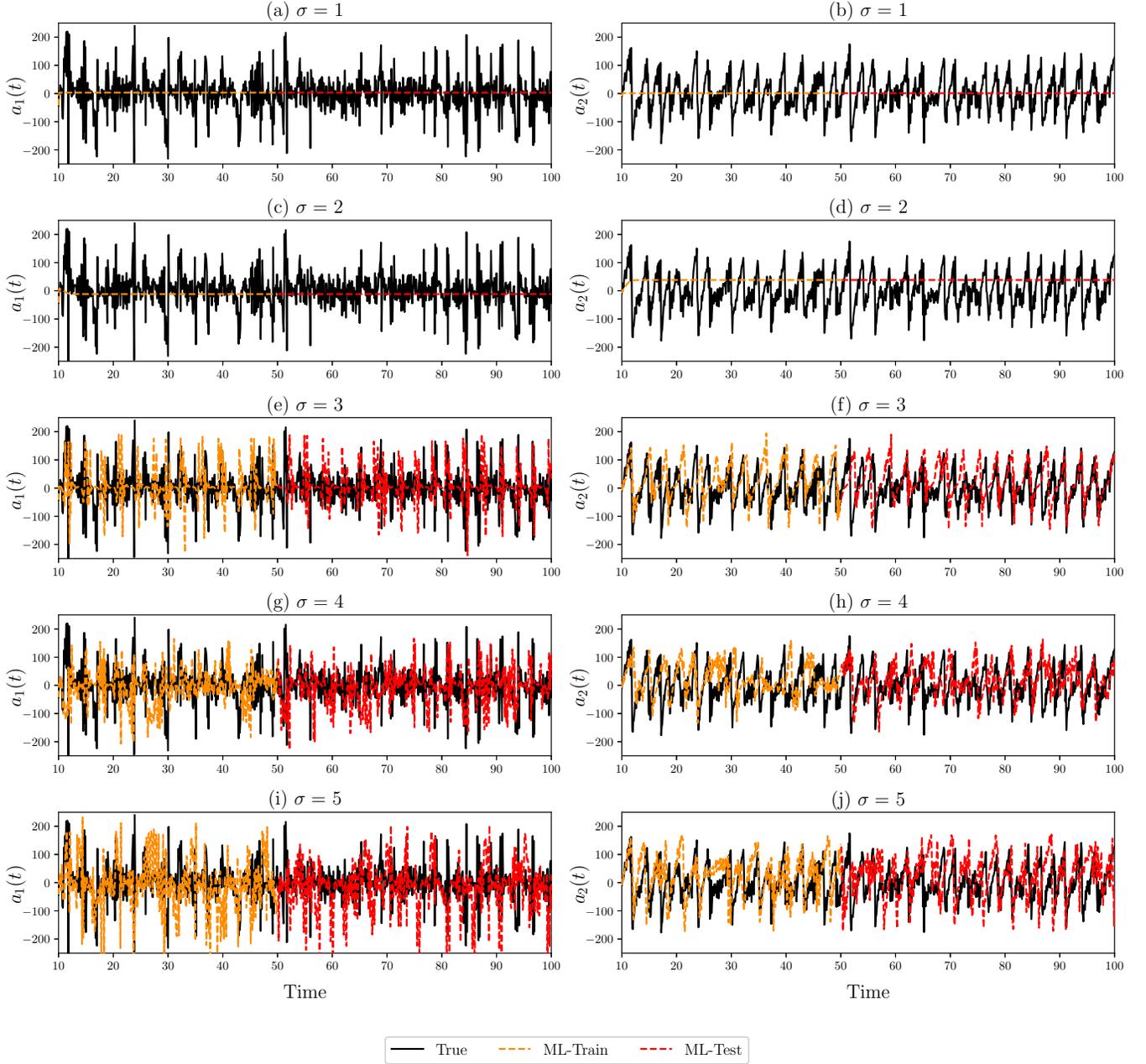}
\caption{Time series evolution of the modal coefficients between $t = 10$ to $t = 100$ for ROM-LSTM simulation based on different lookback time-windows, $\sigma$ and LSTM training with $R = 2$ modes at Re $=450$ and Ro $=3.6 \times 10^{-3}$. (a) $a_1 (t)$ with $\sigma= 1$, (b) $a_{2} (t)$ with $\sigma= 1$, (c) $a_1 (t)$ with $\sigma= 2$, (d) $a_{2} (t)$ with $\sigma= 2$, (e) $a_1 (t)$ with $\sigma= 3$, (f) $a_{2} (t)$ with $\sigma= 3$, (g) $a_1 (t)$ with $\sigma= 4$, (h) $a_{2} (t)$ with $\sigma= 4$, (i) $a_1 (t)$ with $\sigma= 5$, (j) $a_{2} (t)$ with $\sigma= 5$. True projection series is underlined in each figure with black straight line. The training zone is shown with orange dashed line (from $t = 10$ to $t = 50$) and the out-of-sample testing zone is shown with red dashed line (from $t = 51$ to $t = 100$) in ROM-LSTM solution series in each figure.}
\label{fig:7}
\end{figure*}

\begin{figure*}[htbp]
\centering
\includegraphics[width=\mywd]{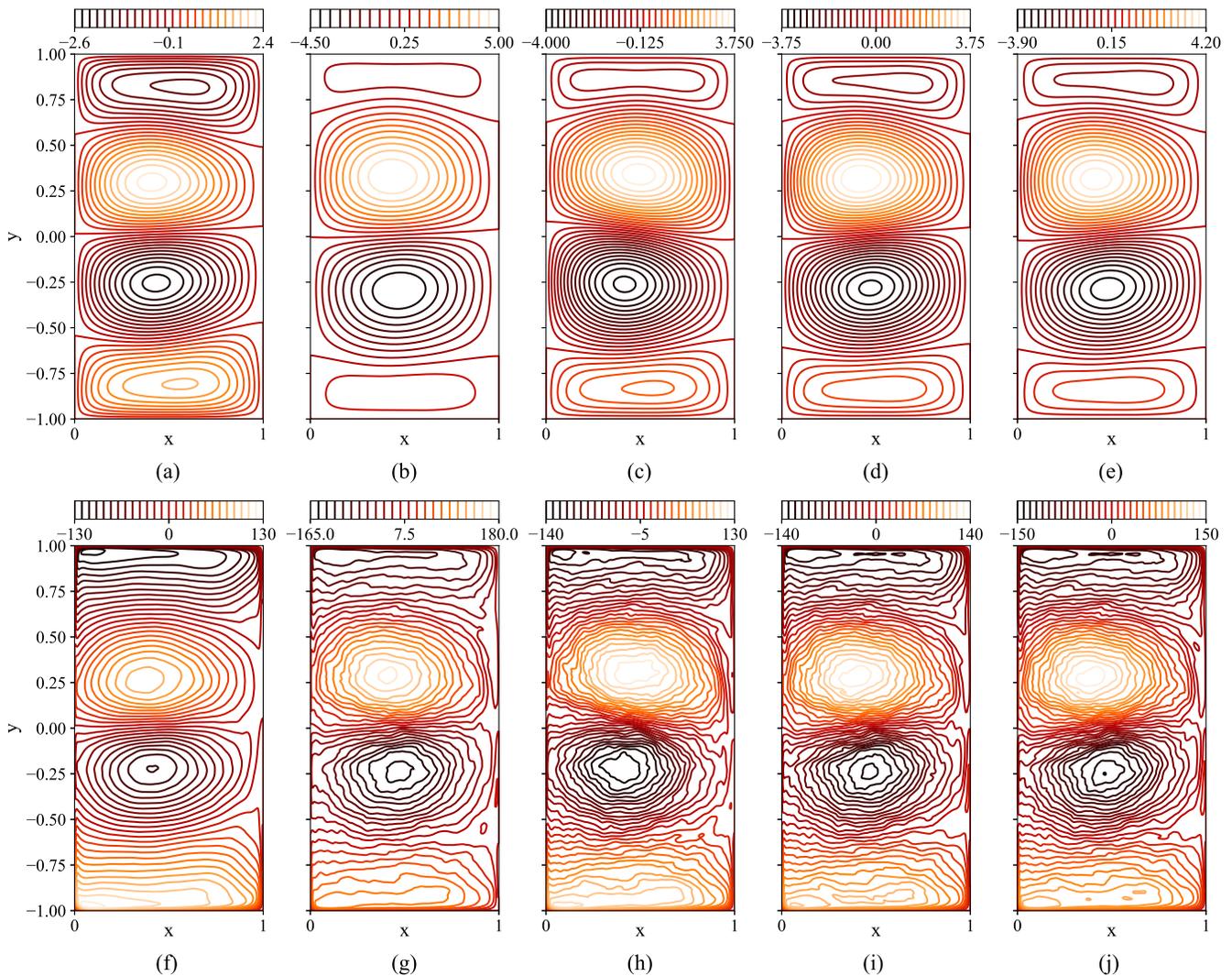}
\caption{Mean streamfunction and vorticity fields obtained by the ROM-LSTM simulation based on different lookback time-windows, $\sigma$ and LSTM training with $R = 2$ modes at Re $=450$ and Ro $=3.6 \times 10^{-3}$. (a) $\psi_{\text{FOM}}$ at a resolution of $256 \times 512$, (b) $\psi_{\text{ROM-LSTM}}$ for LSTM training with $\sigma= 2$, (c) $\psi_{\text{ROM-LSTM}}$ for LSTM training  with $\sigma= 3$, (d) $\psi_{\text{ROM-LSTM}}$ for LSTM training  with $\sigma= 4$, (e) $\psi_{\text{ROM-LSTM}}$ for LSTM training  with $\sigma= 5$, (f) $\omega_{\text{FOM}}$ at a resolution of $256 \times 512$, (g) $\omega_{\text{ROM-LSTM}}$ for LSTM training  with $\sigma= 2$, (h) $\omega_{\text{ROM-LSTM}}$ for LSTM training  with $\sigma= 3$, (i) $\omega_{\text{ROM-LSTM}}$ for LSTM training  with $\sigma= 4$, (j) $\omega_{\text{ROM-LSTM}}$ for LSTM training  with $\sigma= 5$.}
\label{fig:8}
\end{figure*}

Another impressive observation on the predictive capability of the ROM-LSTM framework is presented in Figure~\ref{fig:6} where we show the mean field plots based on the number of modes retained to train the LSTM model. We keep $\sigma = 5$ for this numerical experiment. As we can see the ROM-LSTM model is being able to capture the four-gyre circulation even with only two modes. Indeed, the first few modes contain most of the dynamics in the system and we can also see reduction of some smaller scales for lower mode predictions. Nevertheless, this finding indicates the prediction capability of the ROM-LSTM framework to produce a stable solution of a noisy system. 
However, we have seen the ROM-GP model becomes unstable to predict noisy data set with lower number of modes which makes it very inefficient. In contrast, the proposed non-intrusive framework can be very efficient to produce stable solution with a very few modes. Since we observe promising predictive performance for training with 2 modes only, we present a couple of more analyses on results obtained by the ROM-LSTM framework retaining 2 modes for LSTM training. We can see in Figure~\ref{fig:7} that lower $\sigma$ values simulations are unable to capture the fluctuations along the mean and goes almost straight along the line after a few time states. The model starts to capture the fluctuating flow fields with the increase of $\sigma$ values. The field plots in Figure~\ref{fig:8} also displays the similar conclusions. Since the lower $\sigma$ value solutions stay along the line around the mean (unlike rapid oscillations in ROM-GP solutions), the field plots still show the mean physics to some extent. It is obvious that the model with lower $\sigma$ ignores most of the scales of the system. However, the prediction improves with higher $\sigma$ as displayed in the Figure~\ref{fig:8}.

\begin{figure*}[htbp]
\centering
\includegraphics[width=\mywd]{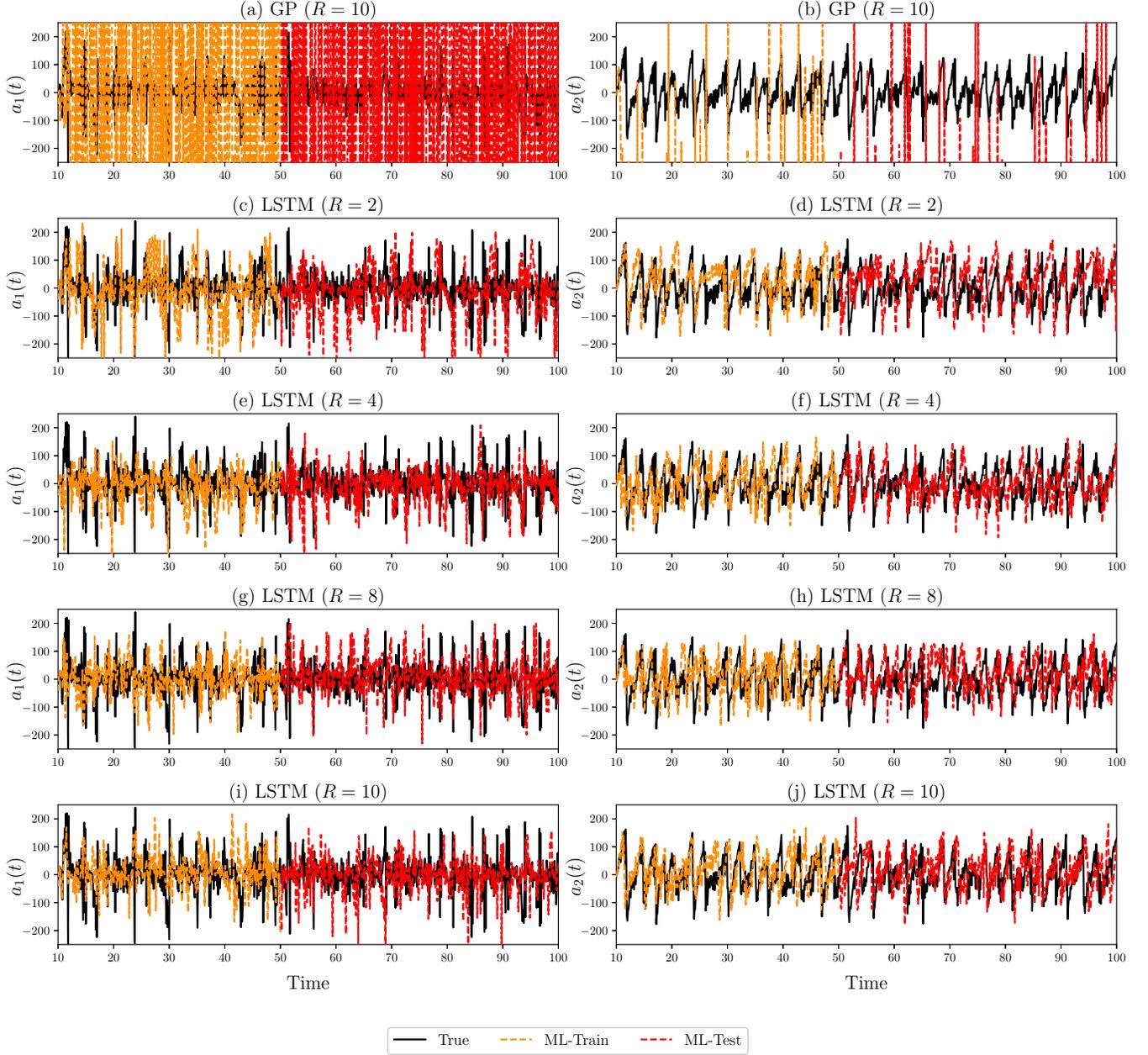}
\caption{Time series evolution of first two modal coefficients, $a_1 (t)$ and  $a_2 (t)$ respectively, between $t = 10$ to $t = 100$ for different ROMs at Re $=450$ and Ro $=3.6 \times 10^{-3}$. (a) $a_1 (t)$ for ROM-GP with $R = 10$ modes, (b) $a_{2} (t)$ for ROM-GP with $R = 10$ modes, (c) $a_1 (t)$ for ROM-LSTM trained with $R = 2$ modes, (d) $a_{2} (t)$ for ROM-LSTM trained with $R = 2$ modes, (e) $a_1 (t)$ for ROM-LSTM trained with $R = 4$ modes, (f) $a_{2} (t)$ for ROM-LSTM trained with $R = 4$ modes, (g) $a_1 (t)$ for ROM-LSTM trained with $R = 8$ modes, (h) $a_{2} (t)$ for ROM-LSTM trained with $R = 8$ modes, (i) $a_1 (t)$ for ROM-LSTM trained with $R = 10$ modes, (j) $a_{2} (t)$ for ROM-LSTM trained with $R = 10$ modes. True projection series is underlined in each figure with black straight line. The training zone is shown with orange dashed line (from $t = 10$ to $t = 50$) and the out-of-sample testing zone is shown with red dashed line (from $t = 51$ to $t = 100$) in ROM-LSTM solution series in each figure.}
\label{fig:9}
\end{figure*}

Finally, we include a comparison plot in Figure~\ref{fig:9} where we present the first two modal coefficients prediction obtained by different ROM set up. The $\sigma$ value is kept 5 for all the ROM-LSTM simulations. As expected, the ROM-GP solutions for 10 modes become totally non-physical and unstable. On the contrary, the ROM-LSTM predictions for $R = 2$, $R = 4$, $R = 8$ and $R = 10$ modes show a good match between the true solution and the prediction. For the quantitative assessment on the accuracy of both ROM-GP and ROM-LSTM frameworks, L\textsubscript{2}-norm errors of the reduced order models (with respect to FOM) for the mean vorticity and streamfunction fields are tabulated in Table~\ref{tab:3}. The root mean-square error or Euclidean L\textsubscript{2}-norm error is computed by:
\begin{align}\label{l2}
   \text{L}\textsubscript{2} = ||e||^2= \sqrt{\frac{1}{N_x N_y}\sum_{i=1}^{N_x}\sum_{j=1}^{N_y} e_{i,j}^2}
\end{align}
where $N_x$ and $N_y$ are the grid resolutions in $x$ and $y$ directions. For the vorticity field, the error i.e., the difference between the predicted mean and FOM solution mean is:
\begin{align}\label{err}
   e_{i,j} = \big|\bar{\omega}_{i,j}^{\text{ROM}} - \bar{\omega}_{i,j}^{\text{FOM}}\big|.
\end{align}

For ROM-LSTM framework, the results are presented for $R = 10$ modes. We can observe that the prediction accuracy increases with the increase in lookback time-window $\sigma$ and we can obtain a more accurate result than the ROM-GP simulation with $R = 80$ using only $10$ modes in ROM-LSTM framework. We present the CPU time per time step (between $t=10$ and $t=100$) for ROM-LSTM framework simulations based on $R = 10$ modes and different $\sigma$ in Table~\ref{tab:4}. We can observe a gradual reduction of computational time (for both training and testing) with lower values of $\sigma$. All the simulations of ROM-LSTM frameworks are done in Python programming environment and CPU time is computed as per time step. The computational time step is set to $1.00 \times 10^{-1}$ for online testing. In our FOM simulation in FORTRAN, the CPU time is about $1.17 \times 10^{-1}$ (between $t=0$ and $t=100$), where computational time step is set $\Delta t=2.5\times 10^{-5}$ due to the CFL restriction of numerical stability for our explicit forward model on the resolution of $256 \times 512$. We refer to Ref.~\cite{rahman2019dynamic} to get an idea about the computational overhead for ROM-GP frameworks using same flow conditions. It should be mentioned that the ROM-GP computations are done using FORTRAN programming platform. Even so, we observe our ROM-LSTM CPU times are in the same order of ROM-GP simulations with $R = 80$ modes. In Table~\ref{tab:4}, we show the computational cost for $R = 10$ modes only since the CPU time (both training and testing) for other ROM-LSTM runs using different modeling conditions are in same order.

\begin{table}[htbp]
\centering
\caption{L\textsubscript{2}-norm errors of the reduced order models (with respect to FOM) for the mean vorticity and streamfunction fields. Note that the ROM-LSTM model trained with $R = 10$ modes results are presented here.}
\begin{tabular}{p{0.22\textwidth}p{0.12\textwidth}p{0.12\textwidth}}
\hline\noalign{\smallskip}
& \ Vorticity & Streamfunction \\
\noalign{\smallskip}\hline\noalign{\smallskip}
\multicolumn{2}{l}{\text{\underline{Intrusive ROM}}} \\
ROM-GP ($R=10$)  & $3.19 \times 10^6$ & \quad $5.59 \times 10^3$  \\
ROM-GP ($R=20$)  & $4.46 \times 10^5$ & \quad $9.87 \times 10^2$  \\
ROM-GP ($R=30$)  & $9.35 \times 10^2$ & \quad $9.99 \times 10^{-1}$  \\
ROM-GP ($R=40$) & $6.60 \times 10^2$ & \quad $4.33 \times 10^{-1}$  \\
ROM-GP ($R=80$) & $1.16 \times 10^3$ & \quad $3.84 \times 10^{-1}$  \\
\multicolumn{2}{l}{\text{\underline{Non-intrusive ROM}}} \\
ROM-LSTM ($\sigma=1$)  & $1.90 \times 10^3$ & \quad $6.12 \times 10^{-1}$  \\
ROM-LSTM ($\sigma=2$)  & $2.50 \times 10^3$ & \quad $7.01 \times 10^{-1}$  \\
ROM-LSTM ($\sigma=3$) & $8.63 \times 10^2$ & \quad $3.65 \times 10^{-1}$ \\
ROM-LSTM ($\sigma=4$)  & $2.69 \times 10^2$ & \quad $2.09 \times 10^{-1}$  \\
ROM-LSTM ($\sigma=5$) & $4.70 \times 10^2$ & \quad $1.29 \times 10^{-1}$   \\
\noalign{\smallskip}\hline
\end{tabular}
\label{tab:3}
\end{table}

\begin{table}[htbp]
\centering
\caption{Computational overhead for the ROM-LSTM model trained with $R = 10$ modes. For training, CPU time is presented as per epoch for 400 samples and for testing, CPU time is presented as per time step. Note that, the time step for testing is set $1.00 \times 10^{-1}$ since the non-intrusive set up is free of numerical stability constraints.}
\begin{tabular}{p{0.14\textwidth}p{0.16\textwidth}p{0.16\textwidth}}
\hline\noalign{\smallskip}
ROM-LSTM & \ Training time (s) & Testing time (s) \\
\noalign{\smallskip}\hline\noalign{\smallskip}
 \quad $\sigma=1$  & \quad \ $8.10 \times 10^{-2}$ & \quad $1.15 \times 10^{-3}$ \\
 \quad $\sigma=2$  & \quad \ $1.07 \times 10^{-1}$ & \quad $1.38 \times 10^{-3}$ \\
 \quad $\sigma=3$ & \quad \ $1.30 \times 10^{-1}$ & \quad $1.56 \times 10^{-3}$ \\
 \quad $\sigma=4$  & \quad \ $1.59 \times 10^{-1}$ & \quad $1.79 \times 10^{-3}$  \\
 \quad $\sigma=5$ & \quad \ $1.80 \times 10^{-1}$ & \quad $2.00 \times 10^{-3}$   \\
\noalign{\smallskip}\hline
\end{tabular}
\label{tab:4}
\end{table}

\section{Summary and Conclusions}
\label{sec:con}
In this paper, we propose an efficient and robust fully non-intrusive ROM framework to capture the large spatio-temporal scale of fluctuating quasi-stationary systems. Due to the robustness and stability of LSTM recurrent neural network in predicting noisy dynamical systems, we consider LSTM architecture to develop our data-driven ROM, denoted as ROM-LSTM in this paper. As an example of large-scale turbulent flows exhibiting a wide range of spatio-temporal scales, we investigate the reduced order modeling of a simple general ocean circulation model, single-layer QG turbulence, to assess the predictive performance of our proposed ROM-LSTM framework. It was previously observed that the conventional physics-based (or intrusive) ROM of QG model requires a large number of POD modes to yield stable and physical flow dynamics. However, the proposed ROM-LSTM framework shows a very promising improvement in reduced order modeling that only a few modes are able to capture a physical solution without any prior knowledge about the underlying governing equations. We first demonstrate that the conventional Galerkin projection ROM approach yields non-physical predictions when we use a small number of representative modes. Although ROM-GP converges to a more physical solution when increasing the number of modes, it does not seem to capture the intermittent bursts appearing in the dynamics of the first few most energetic modes. However, the proposed ROM-LSTM approach is able to capture these bursts and yields remarkably accurate results even when using a small number of modes.

The proposed methodology consists of two phases: offline training and online testing or prediction phase. Initially, we collect the high-fidelity simulation or experimental data snapshots for a certain flow condition. The data snapshots are collected up to a certain time of the full order model simulation for training. Then we do a mapping of the high-resolution instantaneous data snapshots into a reduced order, i.e., low-dimensional space through POD transform. In this process, we generate POD basis functions of the field variables and time dependent modal coefficients for training the LSTM architecture. The LSTM architecture is trained for the modal coefficients based on a preselected lookback time-window, $\sigma$. In the online phase, the trained model is used to predict the modal coefficients recursively for the total time based on initial time history and $\sigma$. Finally, we reconstruct the mean fields for analyses using the predicted coefficients, precomputed basis functions, and mean field values.

We demonstrate the performance of the ROM-LSTM through time series evolution of modal coefficients and mean vorticity and streamfunction fields. To assess the performance of the proposed model, the ROM-LSTM predictions are compared with the high-dimensional solutions as well as with the conventional Galerkin projection based ROM (ROM-GP) solutions. We find that the ROM-LSTM predictions are stable and accurate even with only a couple of POD modes. On the other hand, the ROM-GP framework, as expected, requires a very large number of modes to obtain a physically stable solution, since the ROM-GP framework is susceptible to numerical instability in quasi-stationary flow fields. We further observe that the ROM-LSTM framework gives accurate and physical predictions based on a few time history data points. Indeed, if we increase the value of $\sigma$, the prediction accuracy will increase, but the computational cost of offline training and online prediction will also go up. 
To quantify the accuracy of the prediction of ROM-LSTM framework, we present the L\textsubscript{2}-norm errors for ROM-GP and ROM-LSTM frameworks, which show that the proposed framework trained with 10 modes and $\sigma = 5$ gets a better accuracy than the ROM-GP predictions with 40 or 80 modes.

Based on our findings, we conclude that the ROM-LSTM framework is a better tool than the ROM-GP framework for large-scale quasi-stationary flows in terms of prediction and reduced order modeling. Since the ROM-LSTM framework is fully non-intrusive, it does not rely on the governing equations to obtain the solution, which means that there are no numerical constraints while predicting the solutions. Additionally, it is computationally more efficient to predict the solution using a trained model rather than the physics-based approach of solving ODEs. Hence, the proposed ROM-LSTM framework can be considered a very promising approach in developing a robust and efficient ROMs for large-scale flows with noisy spatio-temporal behavior. In our future works, we will focus on testing the ROM-LSTM framework in more complex three-dimensional turbulent flow problems. We also plan to improve the existing framework based on our findings, and implement the updated framework in suitable ROM applications, such as, flow control, uncertainty quantification, and data assimilation.

\bibliography{reference}
\bibliographystyle{apsrevlong}

\end{document}